\documentclass[twocolumn,pre,aps]{revtex4}

\newcommand{\be}{\begin{equation}} \newcommand{\ee}{\end{equation}} 
\newcommand{\bea}{\begin{eqnarray}}\newcommand{\eea}{\end{eqnarray}}

\usepackage{graphicx}
\begin{document}
\title{ Level statistics of a pseudo-Hermitian Dicke model}
\author{ Tetsuo Deguchi$^{1}$} \email{deguchi@phys.ocha.ac.jp}
\author{ Pijush K. Ghosh$^{2}$} \email{pijushkanti.ghosh@visva-bharati.ac.in}
\author{ Kazue Kudo$^{3}$} \email{kudo.kazue@ocha.ac.jp}
\affiliation{ $^1$ Department of Physics, Graduate School of Humanities and
Sciences, Ochanomizu University,\\
2-1-1 Ohtsuka, Bunkyo-ku, Tokyo 112-8610, Japan}
\affiliation{$^2$ Department of Physics, Siksha-Bhavana,
Visva-Bharati University,\\
Santiniketan, PIN 731 235, India.}
\affiliation{ $^3$ Ochadai Academic Production, Division of Advanced
Sciences, Ochanomizu University, \\
2-1-1 Ohtsuka, Bunkyo-ku, Tokyo 112-8610, Japan}
\begin{abstract} 
A non-Hermitian operator that is related to its adjoint
through a similarity transformation is defined as a pseudo-Hermitian
operator. We study the level-statistics of a pseudo-Hermitian Dicke
Hamiltonian that undergoes Quantum Phase Transition (QPT). We find that
the level-spacing distribution of this Hamiltonian near the integrable
limit is close to Poisson distribution, while it is Wigner distribution
for the ranges of the parameters for which the Hamiltonian is non-integrable.
We show that the assertion in the context of the standard Dicke model
that QPT is a precursor to a change in the 
level statistics is not valid in general.
\end{abstract}
\pacs{05.45.Mt, 64.70.Tg, 02.10.Yn, 03.65.-w}

\maketitle

The study on statistics of energy-levels in quantum many-body systems
has a long history~\cite{mlm,hakke}.
The statistical analysis based on Random Matrix Theory (RMT) has been
applied to characterize quantum chaos and to investigate the integrability
of a quantum system. In particular, it has been conjectured\cite{bgs} that
the level-spacing distribution of an integrable Hermitian Hamiltonian should
be described by the Poisson distribution: $P_{\rm P}(s)=\exp (-s)$.
On the other hand, if the system is non-integrable, the
level-spacing distribution of the Hermitian Hamiltonian should be
given by the Wigner distribution,
i.e., the Wigner surmise for the Gaussian Orthogonal Ensemble (GOE):
$P_{\rm W}(s)=(\pi s/2)\exp(-\pi s^2/4)$.
Although there is no rigorous proof of the Bohigas-Giannoni-Schimdt(BGS)
conjecture\cite{bgs} for
quantum systems, it has been numerically confirmed for a variety of many-body
Hamiltonian~\cite{hakke} and also 
theoretically in the semi-classical limit~\cite{Heusler}.

The RMT without the constraint of Hermiticity was introduced by
Ginibre~\cite{gini} and currently, is an active field of 
research~\cite{akk}. The non-Hermitian RMT
exhibits generic statistical behavior of quantized dissipative systems.
The integrable case corresponds to the Poisson process on the plane,
while a cubic repulsion is a signature of quantum chaotic
scattering~\cite{akk}. An interesting result due to Ginibre~\cite{gini}
is that the probability density function for the eigenvalues of real
Gaussian random non-symmetric matrices with
all the eigenvalues being real is identical to the GOE and consequently, the
level-spacing distribution is given by $P_{\rm W}(s)$. If all the
eigenvalues of a real non-symmetric matrix $M$ are real, it can be shown that
the same matrix can be mapped to its transpose through a similarity
transformation. In particular, $M^T = (X X^T)^{-1} M (X X^T)$, where
the real matrix $X$ diagonalizes $M$ with entirely real eigenvalues $E$,
i.e., $ M= X E X^{-1}$. This shows that Ginibre's ensemble of real
non-symmetric matrices belong to the class of operators known as
pseudo-Hermitian operator, i.e., an operator that is related to its
adjoint through a similarity transformation.

The study on pseudo-Hermitian operators have received considerable attention
recently in connection  with the pioneering work of Bender and
Boettcher~\cite{bend} showing that non-Hermitian operators with
unbroken ${\cal{PT}}$-symmetry admit entirely real spectra. This has opened
up several new directions~\cite{ali,quasi,ddt,ghosh,ghosh1} in the study of
pseudo-Hermitian operators. One of the significant developments is the
construction of pseudo-Hermitian RMT with pseudo-unitary symmetry~\cite{jain}.
The degree of level repulsion is 
different from that of previously known Gaussian orthogonal, unitary and
symplectic ensembles and it seems to point out a new universality class.
Moreover, unlike the Ginibre ensembles (except for the exceptional
case discussed above), non-Hermitian RMT with pseudo-unitary symmetry describes
non-dissipative systems.

In spite of all the above developments, a criterion to characterize
quantum chaos 
and to investigate the integrability of a pseudo-Hermitian operator using
the level-statistics based on RMT is still lacking. Pseudo-Hermitian operators
with entirely real spectra can be shown to be Hermitian with respect to some
modified inner product in the Hilbert space~\cite{ali}. The effect of the
modified inner product in the Hilbert space is to have a modified symplectic
structure for the corresponding classical system. It may be noted here that
a fixed modified inner product in the Hilbert space or the corresponding
symplectic structure is not universal for pseudo-Hermitian systems. It 
varies from one system to another and in general, it is a difficult problem
to identify the proper inner product for a given pseudo-Hermitian system
admitting an entirely real spectrum. It is thus not clear a priory whether the
standard semi-classical analysis in support of the BGS conjecture for
quantum systems with standard inner product in the Hilbert space will
remain unchanged for generic pseudo-Hermitian systems or not. In the absence
of any theoretical support for the validity of the BGS conjecture for
pseudo-Hermitian systems, it may be worth looking for numerical evidences.

The purpose of this paper is to present numerical evidence to show 
that the level-spacing distribution of a non-integrable pseudo-Hermitian
Dicke Hamiltonian with an entirely real spectrum is described by the Wigner
distribution. On the other hand, it approaches to the Poisson
distribution for the parameters of the model close to the integrable limit. 

We consider a pseudo-Hermitian Dicke Hamiltonian (DH) that has been shown
recently to undergo QPT~\cite{ghosh1},
\bea
H & = & \omega \ a^{\dagger} a + 
\omega_0 \ J_z + \frac{\alpha}{\sqrt{2j}} \ e^{i \xi_1} J_-  a^{\dagger} +
\frac{\beta}{\sqrt{2j}}  e^{- i \xi_1} \ J_+ a\nonumber \\
& + & \frac{\gamma}{\sqrt{2j}}  e^{i \xi_2}  J_- a + \frac{\delta}{\sqrt{2j}}
e^{-i \xi_2}  J_+ a^{\dagger},
\label{eq1}
\eea
\noindent where $\omega, \omega_0, 
\alpha, \beta, \gamma, \delta, \xi_1, \xi_2$ are real parameters
and $j$ is the total spin-angular momentum. The operators
$a$, $a^{\dagger}$ are the standard bosonic annihilation-creation
operators and $J_{z}$, $J_{\pm}$ are the generators of the $SU(2)$ algebra,
\bea
&& \left [ a, a^{\dagger} \right ] =1, \nonumber \\
&& \left [ J_+, J_- \right ] = 2 J_z, \ \ \ \ \left [ J_z, J_{\pm} \right ]
\ = \pm J_{\pm}.
\eea
\noindent The Hamiltonian $H$ commutes with the parity operator $\Pi$,
\be
\Pi = e^{i \pi \hat{N}}, \ \ \hat{N} = a^{\dagger} a + J_z + j.
\ee
\noindent The eigenstates of $H$ have definite parity depending on
whether the eigenvalues of the operator $\hat{N}$ are odd or even.
The Hamiltonian (\ref{eq1}) reduces to the standard DH for
$\xi_1=\xi_2=0$ and $\alpha=\beta=\gamma=\delta$. The DH
has been studied extensively from the viewpoint of
QPT~\cite{lieb,hillery,emary}, level-statistics~\cite{emary},
quantum entanglement~\cite{lambert,buzek}, exact
solvability~\cite{hikami} and two-dimensional semiconductor
physics~\cite{datta}.  

The non-Hermitian Hamiltonian $H$ can be mapped to a Hermitian
Hamiltonian ${\cal{H}}= \rho H \rho^{-1}$ through a similarity
transformation when the following relation is satisfied\cite{ghosh1},
\be
\alpha \ \delta - \beta \ \gamma =0.
\label{pseudocon}
\ee
\noindent The operator $\rho$ and ${\cal{H}}$ have
the following forms\cite{ghosh1}:
\bea
\rho & = & \exp \left [ \frac{1}{4} \ln \left ( \frac{\alpha \gamma}
{\beta \delta} \right ) \ \ \left ( J_z + j \right ) \right ], \ \
\frac{\alpha}{\beta} > 0, \ \
\frac{\gamma}{\delta} > 0,\nonumber \\
{\cal{H}} &=& \omega a^{\dagger} a  
+ \omega_0 J_z + \sqrt{ \frac{\alpha \beta}{2j}}
\left ( e^{i \xi_1} \ J_-  a^{\dagger}
+ e^{-i \xi_1} \ J_+ a \right )\nonumber \\
&& + \sqrt{\frac{\gamma \delta}{2j}}
\left ( e^{i \xi_2} \ J_- a +  e^{-i \xi_2} \ J_+ a^{\dagger} \right ).
\label{original}
\eea
\noindent
The Hamiltonian $H$ that is non-Hermitian under the Dirac-Hermiticity
condition becomes Hermitian with respect to the modified inner-product
defined in the Hilbert space as,
$\langle \langle u,v\rangle \rangle_{\eta_+} := \langle u,\eta_+ v \rangle$,
where the metric $\eta_+:= \rho^2$.
In particular,
\be
\langle u | H v \rangle \neq \langle H u | v \rangle, \ \
\langle \langle u|H v \rangle \rangle_{\eta_+} = \langle \langle H u | v
\rangle \rangle_{\eta_+}.
\ee
\noindent Thus, with the modified inner-product, the results of a Hermitian
Hamiltonian follow automatically. Note that
$\langle u | {\cal{H}} v \rangle = \langle {\cal{H}} u | v \rangle$.
We refer to Ref. \cite{ghosh1} for further
details.

In this system, when $j$ is finite, the parity $\Pi$ is a good quantum
number. Two states 
with different parity do not interact with each other. In other words,
we can concentrate on the states with either positive or negative
$\Pi$. Here, we consider the positive-parity states.

The level-spacing distributions are given by
the probability function $P(s)$ of nearest-neighbor spacings 
$s_i=x_{i+1}-x_i$, where $x_i$ are unfolded eigenvalues.
In order to characterize the level-spacing distribution, we employ the
quantity, 
\begin{equation}
 \eta\equiv \left|
\frac{\int_0^{s_0} [P(s)-P_{\rm W}(s)] ds}{\int_0^{s_0} 
 [P_{\rm P}(s)-P_{\rm W}(s)] ds} \right|,
\label{eta}
\end{equation}
where $s_0=0.4729\ldots$ is the intersection point of $P_{\rm P}(s)$ and 
$P_{\rm W}(s)$. We have $\eta=1$ when $P(s)=P_{\rm P}(s)$, and $\eta=0$
when $P(s)=P_{\rm W}(s)$.

In the following, we set $\omega=\omega_0=1$ and $\xi_1=\xi_2=0$ for
convenience. A further choice of $\gamma=\frac{\alpha}{n}$
essentially fixes $\delta$ as $\delta=\frac{\beta}{n}$
due to the pseudo-Hermiticity condition (\ref{pseudocon}), where
$n(\neq -1)$ is a real number. With this parametrization, the
Hamiltonian $H$ can be re-written as,
\be
H=a^{\dagger}a + J_z + \frac{1}{\sqrt{2j}} \left ( \alpha
 J_{-}a^{\dagger}
 + \beta J_{+}a
  +  \frac{\alpha}{n} J_{-}a
 + \frac{\beta}{n} J_{+}a^{\dagger} \right ).
\label{modelH}
\ee
\noindent 
The equivalent Hermitian Hamiltonian ${\cal{H}}$ has the following form:
\be
{\cal{H}} = a^{\dagger}a + J_z +
\sqrt{\frac{\alpha \beta}{2j}} \left [ J_{-}a^{\dagger}
 + J_{+}a + \frac{1}{n} \left ( J_{-}a
 +  J_{+}a^{\dagger} \right ) \right ].
\label{ehH}
\ee
\noindent The total spin-angular momentum $j$ should be large enough to obtain
proper results of level statistics. If $j$ is very small ($j\sim 1$),
because of a kind of finite size effects, level statistics shows no
universal ensembles~\cite{emary}. In our numerical
calculation, $j=10$ unless specifically mentioned.

\begin{figure}
\includegraphics[width=6cm]{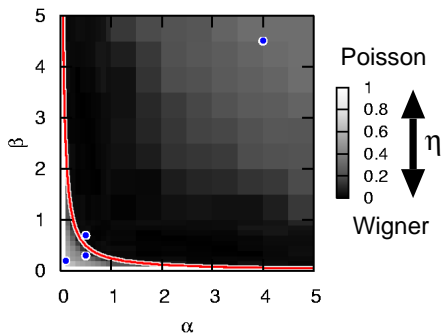}
\caption{\label{Fig1} (color online) Phase diagram of $\eta$, which is
 defined by Eq.~(\ref{eta}), in a special case
 where $\alpha=\gamma$ and $\beta=\delta$. The solid curve corresponds
 to the critical line $\alpha\beta=1/4$. Solid circles correspond to
 the level-spacing distributions in Fig.~\ref{Fig2}.
}
\end{figure}

\begin{figure}
\includegraphics[width=8cm]{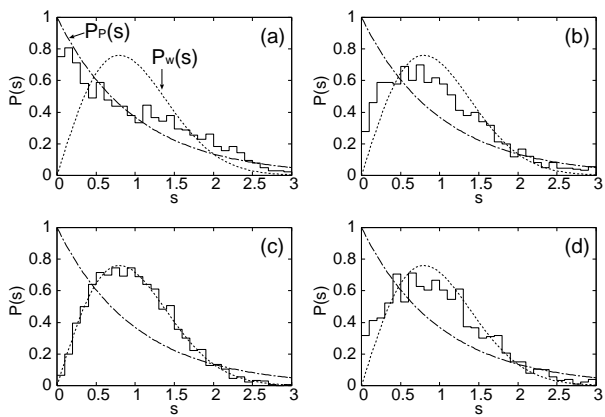}
\caption{\label{Fig2} Level-spacing distributions for the Hamiltonian
 (\ref{modelH}) with $n=1$. (a) Almost Poisson distribution at 
$\alpha=\gamma=0.1$ and $\beta=\delta=0.2$, (b) intermediate
 distribution at $\alpha=\gamma=0.5$ and 
$\beta=\delta=0.3$, (c) Wigner distribution at $\alpha=\gamma=0.5$ and
 $\beta=\delta=0.7$, and 
(d) deformed Wigner distribution at $\alpha=\gamma=4$ and $\beta=\delta=4.5$. 
}
\end{figure}

Figure~\ref{Fig1} exhibits the phase diagram of $\eta$ for $H$ in
Eq. (\ref{modelH}) with $n=1$. 
At the critical line $\alpha\beta=1/4$, $\eta$ rapidly changes. 
For $\alpha\beta < 1/4$, level
statistics is almost Poissonian as seen in Fig.~\ref{Fig2}(a) 
\cite{integrable-case}. 
As $\alpha\beta$ increases, $P(s)$ changes from the Poisson to Wigner
distributions: It gives an intermediate distribution,
e.g. Fig.~\ref{Fig2}(b).  
For $\alpha\beta > 1/4$, level-spacing distribution is
almost given by the Wigner distribution, as shown in
Fig.~\ref{Fig2}(c). However, as $\alpha\beta$ increases further, 
$\eta$ gradually increases. 
In other words, level statistics gradually
changes from the Wigner distribution to Poissonian one again as $\alpha\beta$
becomes very large. Figure~\ref{Fig2}(d) is an example of an intermediate
distribution for large $\alpha\beta$.   
The behavior of $\eta$ along the line $\alpha=\beta$ in Fig.~\ref{Fig1}
corresponds to that of Ref.~\cite{emary}.

The Hamiltonian ${\cal{H}}$ in Eq. (\ref{ehH}) corresponds to the standard
DH for $n=1$ and has been studied in some detail in Ref.~\cite{emary}.
Based on the results of \cite{emary} and the numerical findings for $H$
as described above, we suggest the following:\\
(i) The criteria to distinguish
between integrable and non-integrable phases of a Hermitian Hamiltonian
are valid also for a pseudo-Hermitian Hamiltonian.
The quantum Hamiltonian ${\cal{H}}$ and $H$ have the identical eigenvalues,
since they are related to each other through a
similarity transformation. 
Thus, both $H$ and ${\cal{H}}$ show similar changes in the level-spacing
distributions as a function of $\alpha \beta$.\\
(ii) The onset of quantum chaos in a pseudo-Hermitian Hamiltonian is
manifested by a change in the level statistics from Poissonian to Wigner
distribution. 
It is known that the semi-classical Hermitian Hamiltonian corresponding to
${\cal{H}}$ shows chaotic behavior for $\alpha \beta > \frac{1}{4}$
and regular periodic orbits are obtained for
$\alpha \beta < \frac{1}{4}$~\cite{emary}. It is also the case for the
Hamiltonian $H$ in the semi-classical limit. In fact, a non-Hermitian
Hamiltonian and its equivalent Hermitian Hamiltonian describe the same
physics in the classical limit within the formalism of pseudo-Hermitian
quantum physics and the correspondence principle~\cite{ali}. \\

\begin{figure}
\includegraphics[width=7.5cm]{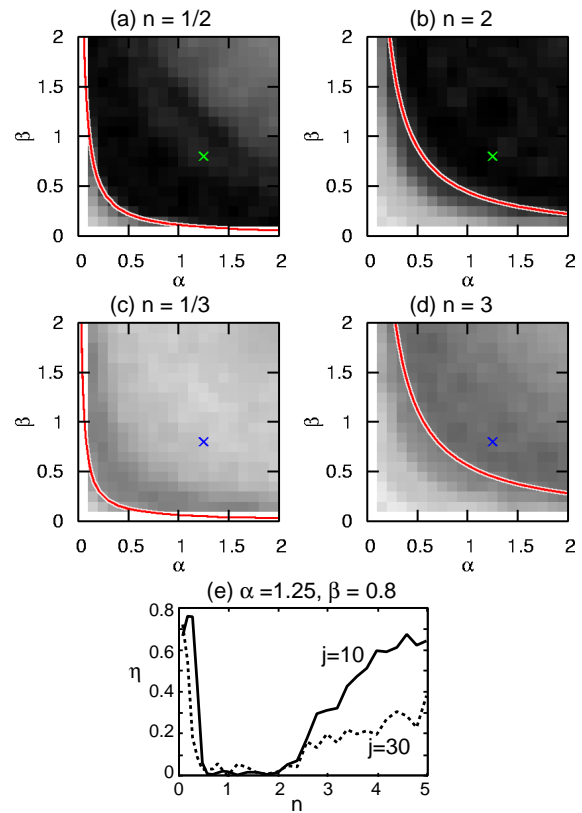}
\caption{\label{Fig3} (color online) Phase diagram of $\eta$, which is
 defined by Eq.~(\ref{eta}) for $\gamma = \alpha /n$ and
$\delta =\beta /n$; (a) $n=1/2$, (b) $n=2$, (c) $n=1/3$, and (d) $n=3$.
The solid curve in each panel is given by $\alpha\beta=n^2/(n+1)^2$.
(e) Dependence of $\eta$ plotted against $n$ for $\alpha=1.25$ and
$\beta=0.8$. The parameter combination is shown in (a) -- (d) by
the points ($\times$).
}
\end{figure}
We now present numerical results for other values of $n$.
Figures~\ref{Fig3}(a) and \ref{Fig3}(b) are phase-diagrams of $\eta$
for $n=\frac{1}{2}$ and $n=2$, respectively. Changes in
level-statistics appear around the critical-line determined by
$\alpha\beta=\frac{n^2}{(n+1)^2}$
for both the cases, i.e., the physical picture is identical to the case of
$n=1$.
However, the change of $\eta$ around the critical line is small in
Figs.~\ref{Fig3}(c) and 
\ref{Fig3}(d), which exhibit phase-diagrams of $\eta$ for $n=\frac{1}{3}$
and $n=3$, respectively. In fact, level statistics does not show clear
Wigner behavior for $\alpha\beta > \frac{n^2}{(n+1)^2}$ with
$n=\frac{1}{3}$ and $n=3$.

In order to see the change of $\eta$ clearly, we depict $\eta$ as a
function of $n$ for $\alpha\beta=1$ in Fig.~\ref{Fig3}(e), where the
curve of $\eta$ for $j=10$ is compared with that for $j=30$. 
The non-zero $\eta$ behavior for $n > 2$ and $n < \frac{1}{2}$ in
Fig.~\ref{Fig3}(e) indicates that Poisson behavior continues beyond the
critical line in the phase diagram of $\eta$ for those
regimes of $n$. The behavior is a manifestation
of the fact that the system becomes close to the integrable limit,
if $|n| \ll 1$ or $|n| \gg 1$. One plausible explanation to understand
the origin of the precise critical values of $n$ (i.e., $\frac{1}{2}$ and $2$)
may lie in the non-applicability of perturbation techniques by treating
either the counter-rotating ( for $ 1 < n \leq 2$ ) or the rotating terms
( for $ 1 > n \geq \frac{1}{2}$) as perturbation.


A comment is in order at this point.
The pseudo-Hermitian Dicke model is known to undergo QPT with the critical-line
determined by the equation $\alpha\beta = \frac{n^2}{(n+1)^2}$~\cite{ghosh1}.
We suggest that the value of $\eta$, which characterizes the level statistics,
is a possible measure to estimate the onset of the QPT not only for the
Hermitian Hamiltonian ${\cal{H}}$, but, also for the quasi-Hermitian
Hamiltonian $H$. However, the usual assertion \cite{emary} within the
context of the standard Dicke model that QPT is a precursor to a change
in the level statistics is not valid in general for the pseudo-Hermitian
Dicke model which has a larger parameter space.
According to the relevant assertion in Ref. \cite{emary}, 
level statistics is given by the Wigner 
distribution for $\alpha\beta > \frac{n^2}{(n+1)^2}$. For any positive
$n$, the inequality is satisfied when $\alpha\beta=1$. Therefore, if the
assertion were always valid, $\eta$ would be always zero (or very
small) for $n>0$, which is certainly not the case for $n > 2$ and
$n < \frac{1}{2}$ in Fig.~\ref{Fig3}(e). 
The assertion is valid for the pseudo-Hermitian Dicke
model for $ \frac{1}{2} < n < 2$ of which the standard Dicke model
corresponding to $n=1$ appears as a special case.

We conclude with the following:\\
(i) Based on our numerical results, we conjecture that 
the level-spacing distribution comes close to the Poisson distribution
$P_{\rm P}(s)$ as the system approaches the integrable limit, 
while for the non-integrable pseudo-Hermitian Hamiltonian it should be
described by the Wigner distribution $P_{\rm W}(s)$.\\
(ii) We have also shown that the assertion \cite{emary} that QPT is
a precursor to a change in the level statistics in the standard Dicke model
is not valid in general for the pseudo-Hermitian Dicke model which has
a larger parameter space. The assertion holds true for the pseudo-Hermitian
Dicke model only for a limited range of the parameter space.

\textit{Acknowledgments.}
PKG would like to thank Ochanomizu University for
warm hospitality during his visit under the JSPS Invitation
Fellowship for Research in Japan(S-08042), where a part of this work
has been carried out.


\begin{thebibliography}{99}
\bibitem{integrable-case}
Fig.~\ref{Fig2}(a) is not a perfect Poisson
distribution. When $\alpha=\beta=\gamma=\delta=0$, 
the system is integrable, but the level statistics is non-Poissonian:
All the nonzero level spacings are identical and 
the eigenvalues are highly degenerated.  
For small nonzero $\alpha\beta$, level statistics is influenced by the
property.
\end{thebibliography}
\end{document}